\documentstyle[12pt,epsf]{article} 

\def\lsim{\mathrel{\lower2.5pt\vbox{\lineskip=0pt\baselineskip=0pt 
           \hbox{$<$}\hbox{$\sim$}}}} 
\def\gsim{\mathrel{\lower2.5pt\vbox{\lineskip=0pt\baselineskip=0pt 
           \hbox{$>$}\hbox{$\sim$}}}} 
\def\ad{\dot{a}}
\def\bd{\dot{b}}
\def\nd{\dot{n}}
\def\cd{\dot{c}}
\def\add{\ddot{a}}
\def\bdd{\ddot{b}}

\def\cdd{\ddot{c}}
\def\ap{a^{\prime}}
\def\bp{b^{\prime}}
\def\np{n^{\prime}}
\def\cp{c^{\prime}}
\def\Ap{A^{\prime}}
\def\Cp{C^{\prime}}
\def\app{a^{\prime\prime}}

\def\npp{n^{\prime\prime}}
\def\cpp{c^{\prime\prime}}
\def\App{A^{\prime\prime}}
\def\Cpp{C^{\prime\prime}}

\def\k{\kappa}

\def\d{\delta}

\def\L{\Lambda}
\def\e{{\rm exp}}
\def\l{\lambda}
\begin{document} 
\begin{flushright}
DPNU-01-08\\ hep-th/0105186
\end{flushright}

\vspace{10mm}

\begin{center}
{\Large \bf 
Warped Geometry in Higher Dimensions with\\ an Orbifold Extra Dimension}

\vspace{20mm}
Masato ITO 
\footnote{E-mail address: mito@eken.phys.nagoya-u.ac.jp}
\end{center}

\begin{center}
{
\it 
{}Department of Physics, Nagoya University, Nagoya, 
JAPAN 464-8602 
}
\end{center}

\vspace{25mm}

\begin{abstract}
 We solve the Einstein equations in higher dimensions with warped
 geometry where an extra dimension is assumed to have orbifold symmetry
 $S^{1}/Z_{2}$. 
 The setup considered here is an extension of the five-dimensional 
 Randall-Sundrum model to $5+D$ dimensions, and
 hidden and observable branes are fixed on the orbifold.
 It is assumed that the brane tension (self-energy) of each brane with 
 $(4+D)$-dimensional spacetime is anisotropic 
 and that the warped metric function of the four dimensions is 
 generally different from that of the extra $D$ dimensions.
 We point out that the forms of the warped metric functions and the relations 
 between the tensions of two branes depend on the
 integration constant appearing in the Einstein equations 
 as well as on the sign of the bulk cosmological constant.
\end{abstract} 
\newpage 
%%%%%%%%%%%%%%%%%%%%%%%%%%%%%%% Intro %%%%%%%%%%%%%%%%%%%%%%%%%%%%%%%%
\section{Introduction}

 $\hspace{0.5cm}$
 Motivated by the Ho$\check{\rm r}$ava-Witten model in 11-dimensional theory 
 ($M$ theory) compactified on the orbifold $S^{1}/Z_{2}$, many
 models have been proposed using the notion that there are
 two branes that represent the boundaries of higher dimensional
 spacetime \cite{HW}.
 Consequently, there has been growing interest among particle physicists
 and cosmologists in models with extra dimensions.
 Recent developments are based on the idea that ordinary
 matter fields could be confined to a three-brane world embedded in the
 higher dimensional space.

 Adopting this idea further, 
 there are several proposals that try to
 relate the Planck scale of the observable world to the higher
 dimensional Planck scale. 
 In the model proposed by Arkani-Hamed Dimopoulos and Dvali \cite{LE},
 the fundamental scale $M_{\ast}$ can be related to the usual 
 four-dimensional Planck scale $M_{p}$ via a volume factor,
 $M^{2}_{p}=M^{n+2}_{\ast}R^{n}$, where $R^{n}$ is the volume of
 the compact space and $n$ is the number of extra dimensions. 
 If $R$ is sufficiently large, $M_{\ast}$ can be as low as the
 $1\;TeV$ scale; thus the model gives a possible solution to 
 the gauge hierarchy problem.
 Furthermore, Randall and Sundrum \cite{Randall:1999ee,Randall:1999vf} have
 presented a static solution to the classical five-dimensional 
 Einstein equations with negative bulk cosmological constant 
 (AdS space).
 The warped metric (factor) in the model is an exponential scaling of 
 the metric along the fifth dimension compactified on the 
 $S^{1}/Z_{2}$ orbifold.
 This solution appeals to the possibility of an extra
 dimension limited by two three-branes with opposite tensions, and
 provides an alternative explanation for the hierarchy problem as due to
 the warped factor if the observable brane has negative brane tension.
 Both approaches assume that the standard model particles are
 confined to a three-brane embedded in higher dimensional spacetime
 and that gravity exists in the bulk.

 An important question concerning these kinds of model is
 whether or not standard four-dimensional gravity is reproduced on the brane.
 In the Randall-Sundrum model, even if the fifth dimension is uncompactified,
 the usual gravity is shown to be recovered because of the 
 existence of a massless graviton trapped in the brane \cite{Randall:1999vf}. 
 Further, another problem is that the stabilization mechanism
 for the size of the extra dimensions is yet unknown.
 Introducing a bulk scalar field that interacts with the branes, 
 several mechanisms have been proposed \cite{RSstabi}.
 On the other hand, the existence of the 
 extra dimensions allows much phenomenology, including 
 the production of Kaluza-Klein excitations of gravitons at future 
 colliders or their detection in high precision measurements 
 at low energies.

 The Randall-Sundrum static solution has been extended to time
 dependent solutions and their cosmological properties have been
 extensively studied \cite{RScosm}.
 In the framework of brane world cosmology,
 the serious problem emphasized recently is an unusual form of the
 Friedmann equations in the case of one extra dimension, which
 leads to a particular behavior of the Hubble parameter on the brane.
 In particular, the Hubble parameter $H$ is proportional to the energy
 density on the brane instead of the familiar dependence
 $H\sim\sqrt{\rho}$.

 The purpose of this paper is to extend the five-dimensional Randall-Sundrum
 model to higher dimensional cases.
 Since the original Randall-Sundrum model was inspired by superstring
 theory or $M$ theory,
 the version in higher dimensions should be naturally motivated.
 In this case, we are interested in whether the tension of the higher 
 dimensional brane is anisotropic or not, 
 and in the relation between the brane tension and the bulk cosmological 
 constant.
 We study the metric of the $(5+D)$-dimensional Randall-Sundrum model, 
 where the $5+D$ dimensions are composed of the $(4+D)$-dimensional 
 spacetime and a dimension compactified on the orbifold $S^{1}/Z_{2}$,
 $|y|\leq L$.
 The $(4+D)$-dimensional world resides in the $(3+D)$-brane, and
 two branes are fixed at $y=0$ and $y=L$.
 The observable brane we live in is assumed to be the brane at $y=L$
 and the hidden brane is at $y=0$.
 The ways of taking the metric ansatz are various. 
 In this paper, we consider the case that the four-dimensional warped
 metric function $a(y)$ is generally different from 
 the extra $D$-dimensional warped metric function $c(y)$.
 Recently, the scenario in which  $a(y)$ is equal to  $c(y)$ 
 has been discussed \cite{metric}. 
 Furthermore, we assume that
 the brane tension of the $4+D$ dimensions is anisotropic, namely,
 the brane tension of the four-dimensional spacetime is generally
 different from that of the extra $D$-dimensional space. 
 Based on the above assumptions,
 we solve the $(5+D)$-dimensional Einstein equation with the bulk
 cosmological constant and study the forms of $a(y)$ and $c(y)$ explicitly.
 Moreover, we derive the relations between the brane tension of the
 four dimensions and that of the extra $D$ dimensions and represent the
 behavior of each brane tension for the distance between two branes.

 This paper is organized as follows.
 In section 2, the setup considered here is described and
 generalized Einstein equations with time dependence are explicitly expressed.
 In the simplest case of an isolated two-brane system, 
 we give the higher dimensional Friedmann-type equation on the brane.
 In section 3, we solve the static $(5+D)$-dimensional Randall-Sundrum
 model with the bulk cosmological constant $\L$.
 For each case of $\L <0$, $\L >0$, and $\L=0$, the $(4+D)$-dimensional
 metric functions can be obtained. 
 We show that the forms of the warped metric functions and the relation
 between the brane tensions on the orbifold depend on the integration
 constant appearing in the Einstein equations as well as on the sign of the
 bulk cosmological constant.
 A summary and discussion are given in the final section.
 In an appendix, we review the Kasner solution  of
 the $(4+D)$-dimensional anisotropic cosmological model.
%
%%%%%%%%%%%%%%%%%%%%%%%%%%%%%% 2 %%%%%%%%%%%%%%%%%%%%%%%%%%%%%%%%
\section{The Setup}
%%%%%%%%%%%%%%%%%%%%%%%%%%%%%%%%%%%%%%%%%%%%%%%%%%%%%%%%%%%%%%%%%

 $\hspace{0.5cm}$
 We consider the higher dimensional spacetime with
 an orbifold extra dimension.
 This setup is an extension of the Randall-Sundrum model with
 five-dimensional warped metric.
 The two $3+D$ branes with the $(4+D)$-dimensional spacetime embedded
 in the $(5+D)$-dimensional
 spacetime are located at $y=0$ and at $y=L$, where the $y$ direction 
 is compactified on the orbifold $S^{1}/Z_{2}$.
 This $(5+D)$-dimensional model is described by the action
 \begin{eqnarray}
  S=\int^{L}_{-L}dy\int d^{4+D}x\sqrt{|g|}
     \left(\frac{1}{2\k^{2}}{\cal R}-\L\right)\label{eqn1}
 \end{eqnarray}
 in bulk, where $1/\k^{2}$ is the fundamental gravitational scale and
 $\L$ is the bulk cosmological constant.

 To solve the Einstein equations, the metric ansatz can be written
 in the following form
 \begin{eqnarray}
  ds^{2}&=&n^{2}(t,y)dt^{2}-a^{2}(t,y)d\vec{x}^{2}-b^{2}(t,y)dy^{2}
        -c^{2}(t,y)\left(dz^{2}_{1}+\cdots +dz^{2}_{D}\right)\nonumber\\
  &\equiv&g_{AB}dx^{A}dx^{B}\,,
  \label{eqn2}
 \end{eqnarray}
 where $A,B=0,\cdots, 4+D$.
 We shall use the notation $\{x^{\mu}\}$ with $\mu=0,\cdots,3$ for the
 coordinates on the four-dimensional spacetime $\{t,\vec{x}\}$, 
 $x^{4}=y$ for a coordinate on the orbifold compactification, 
 and $\{x^{a}\}$ with $a=5,\cdots,D+4$ for coordinates on the extra 
 $D$-dimensional space $\{z_{1},\cdots,z_{D}\}$.
 It is assumed that the distribution of the brane tension and the matter 
 on the brane with $(4+D)$-dimensional spacetime is anisotropic.
 The Einstein tensor corresponds to
 \begin{eqnarray}
  G_{AB}={\cal R}_{AB}-\frac{1}{2}g_{AB}{\cal R}\,,
  \label{eqn3}
 \end{eqnarray}
 where ${\cal R}_{AB}$ and ${\cal R}$ represent the Ricci tensor and 
 the scalar curvature, respectively.
 The Einstein equation is given by $G_{AB}=\kappa^{2}T_{AB}$,
 where $T_{AB}$ is the energy-momentum tensor.
 It is assumed that there are contributions to $T_{AB}$ from the bulk and
 the branes as
  \begin{eqnarray}
  T_{AB}&=&T^{\rm bulk}_{AB}+T^{\rm brane}_{AB}\,.\label{eqn4}
  \end{eqnarray}
 From the bulk we have
  \begin{eqnarray}
  T^{\rm bulk}_{AB}&=& g_{AB}\Lambda\,,\label{eqn5}
  \end{eqnarray}
 where $\L$ is the cosmological constant in the bulk,
 and from the two branes
  \begin{eqnarray}
  T^{A,\rm brane}_{B}&=&
  \frac{\d(y)}{b}
  diag(\;V_{1}+\rho_{1},V_{1}-p_{1},V_{1}-p_{1},V_{1}-p_{1},0
      ,\underbrace{V^{\ast}_{1}-p^{\ast}_{1},\cdots,
       V^{\ast}_{1}-p^{\ast}_{1}}_{D}\;)\nonumber\\
  &&+\frac{\d(y-L)}{b}
  diag(\;V_{2}+\rho_{2},V_{2}-p_{2},V_{2}-p_{2},V_{2}-p_{2},0,
       \nonumber\\
  &&\hspace{8cm}
      \underbrace{V^{\ast}_{2}-p^{\ast}_{2},\cdots,
       V^{\ast}_{2}-p^{\ast}_{2}}_{D}\;)\,.\nonumber\\\label{eqn6}
 \end{eqnarray}
 Here the indices $1$ and $2$ denote the brane at $y=0$ and at $y=L$,
 respectively. 
 $V$, $\rho$, and $p$ represent the brane tension, the density, and the
 pressure of the matter on each brane, respectively.
 The superscript $\ast$ corresponds to quantities in the extra 
 $D$-dimensional space.
 Using the metric ansatz Eq.(\ref{eqn2}), we can write the 
 Einstein equation for each component.
 The $(0,0)$ component for the $t$ direction is given by
 \begin{eqnarray}
%(0,0)
 &&\frac{1}{n^{2}}
 \left[
 3\left(\frac{\ad}{a}\right)^{2}
 +\frac{1}{2}D(D-1)\left(\frac{\cd}{c}\right)^{2}
 +3\frac{\ad}{a}\frac{\bd}{b}
 +D\frac{\bd}{b}\frac{\cd}{c}
 +3D\frac{\ad}{a}\frac{\cd}{c}\right]\nonumber\\
 &&
 -\frac{1}{b^{2}}
  \left[
  3\left(\frac{\ap}{a}\right)^{2}
 +\frac{1}{2}D(D-1)\left(\frac{\cp}{c}\right)^{2}
 +3\frac{\app}{a}  
  -3\frac{\ap}{a}\frac{\bp}{b}
 +D\frac{\cpp}{c}+3D\frac{\ap}{a}\frac{\cp}{c}
 -D\frac{\bp}{b}\frac{\cp}{c}
  \right]\nonumber\\
 &&=\kappa^{2}\Lambda
 +\kappa^{2}\frac{V_{1}+\rho_{1}}{b}\d(y)
 +\kappa^{2}\frac{V_{2}+\rho_{2}}{b}\d(y-L)\,.\label{eqn7}
 \end{eqnarray}
 The $(i,i)$ component for the three-dimensional space ($i=1,2,3$) is
 \begin{eqnarray}
%(i,i)
 &&\frac{1}{n^{2}}
 \left[
 -2\frac{\add}{a}-\frac{\bdd}{b}-D\frac{\cdd}{c}
 -\left(\frac{\ad}{a}\right)^{2}
 -\frac{1}{2}D(D-1)\left(\frac{\cd}{c}\right)^{2}\right.\nonumber\\
 &&\hspace{5cm}\left.
 -2\frac{\ad}{a}\frac{\bd}{b}
 -2D\frac{\ad}{a}\frac{\cd}{c}
 -D\frac{\bd}{b}\frac{\cd}{c}
 +2\frac{\ad}{a}\frac{\nd}{n}+\frac{\bd}{b}\frac{\nd}{n}
 +D\frac{\cd}{c}\frac{\nd}{n}
 \right]\nonumber\\
 &&+\frac{1}{b^{2}}
    \left[
  2\frac{\app}{a}+\frac{\npp}{n}
      +D\frac{\cpp}{c}
      +\left(\frac{\ap}{a}\right)^{2}
      +\frac{1}{2}D(D-1)\left(\frac{\cp}{c}\right)^{2}\right.\nonumber\\
 &&\hspace{5cm}\left.
  -2\frac{\ap}{a}\frac{\bp}{b}
      +2\frac{\ap}{a}\frac{\np}{n}
      -\frac{\bp}{b}\frac{\np}{n}
      +2D\frac{\ap}{a}\frac{\cp}{c}
      -D\frac{\bp}{b}\frac{\cp}{c}
      +D\frac{\cp}{c}\frac{\np}{n}
    \right]\nonumber\\
 &&=-\kappa^{2}\Lambda
  -\kappa^{2}\frac{V_{1}-p_{1}}{b}\d(y)
  -\kappa^{2}\frac{V_{2}-p_{2}}{b}\d(y-L)\,.\label{eqn8}
 \end{eqnarray}
 For the $(4,4)$ component for the $y$ direction compactified on $S^{1}/Z_{2}$
 we get 
 \begin{eqnarray}
%(4,4)
 &&
 \frac{1}{n^{2}}
 \left[
        -3\frac{\add}{a}-D\frac{\cdd}{c}
        -3\left(\frac{\ad}{a}\right)^{2}
        -\frac{1}{2}D(D-1)\left(\frac{\cd}{c}\right)^{2}
        +3\frac{\ad}{a}\frac{\nd}{n}
        +D\frac{\cd}{c}\frac{\nd}{n}
        -3D\frac{\ad}{a}\frac{\cd}{c}
 \right]\nonumber\\
 &&+\frac{1}{b^{2}}
 \left[
 3\left(\frac{\ap}{a}\right)^{2}
 +\frac{1}{2}D(D-1)\left(\frac{\cp}{c}\right)^{2}
 +3\frac{\ap}{a}\frac{\np}{n}
 +3D\frac{\ap}{a}\frac{\cp}{c}
 +D\frac{\np}{n}\frac{\cp}{c}
 \right]=-\kappa^{2}\Lambda\,.\label{eqn9}
 \end{eqnarray}
 The $(a,a)$ component for the $D$-dimensional space ($a=5,\cdots, D+4$)
 takes the form
 \begin{eqnarray}
%(5,5)
 &&\frac{1}{n^{2}}
 \left[
 -3\frac{\add}{a}-\frac{\bdd}{b}
 -(D-1)\frac{\cdd}{c}
 -3\left(\frac{\ad}{a}\right)^{2}
 -\frac{1}{2}(D-1)(D-2)\left(\frac{\cd}{c}\right)^{2}\right.\nonumber\\
 &&\hspace{2cm}\left.
 -3\frac{\ad}{a}\frac{\bd}{b}
 -(D-1)\frac{\bd}{b}\frac{\cd}{c}
 -3(D-1)\frac{\ad}{a}\frac{\cd}{c}
 +3\frac{\ad}{a}\frac{\nd}{n}
 +\frac{\bd}{b}\frac{\nd}{n}
 +(D-1)\frac{\cd}{c}\frac{\nd}{n}
 \right]\nonumber\\
&&+\frac{1}{b^{2}}
    \left[
          3\frac{\app}{a}+\frac{\npp}{n}
      +(D-1)\frac{\cpp}{c}
      +3\left(\frac{\ap}{a}\right)^{2}
      +\frac{1}{2}(D-1)(D-2)\left(\frac{\cp}{c}\right)^{2}
       \right.\nonumber\\
 &&\hspace{2cm}\left.
      -3\frac{\ap}{a}\frac{\bp}{b}
      +3\frac{\ap}{a}\frac{\np}{n}
      -\frac{\bp}{b}\frac{\np}{n}
      +3(D-1)\frac{\ap}{a}\frac{\cp}{c}
      -(D-1)\frac{\bp}{b}\frac{\cp}{c}
      +(D-1)\frac{\np}{n}\frac{\cp}{c}
    \right]\nonumber\\
 &&=-\kappa^{2}\Lambda
  -\kappa^{2}\frac{V^{\ast}_{1}-p^{\ast}_{1}}{b}\d(y)
  -\kappa^{2}\frac{V^{\ast}_{2}-p^{\ast}_{2}}{b}\d(y-L)\,\label{eqn10}
 \end{eqnarray}
 and the non diagonal $(0,4)$ component for the $t$ and $y$ directions
 is written as
 \begin{eqnarray}
 -3\frac{\dot{\ap}}{a}+3\frac{\np}{n}\frac{\ad}{a}
     +3\frac{\ap}{a}\frac{\bd}{b}+D\frac{\np}{n}\frac{\cd}{c}
     -D\frac{\cd^{\prime}}{c}+D\frac{\bd}{b}\frac{\cp}{c}=0
 \,.\label{eqn11}
 \end{eqnarray}
 Here the primes (overdots) denote the derivatives with respect to $y$ ($t$).
 Although the functions $a,n,$ and $c$ are continuous at the brane, their 
 derivatives with respect to $y$ are discontinuous because of the presence
 of the brane.
 By matching the coefficients of the delta functions,
 the $(0,0)$, $(i,i)$, and $(a,a)$ components of the Einstein equations
 are subject to jump conditions on the first derivatives of the functions.
 In order to derive jump conditions on $a,n,$ and $c$, 
 we define the function \cite{RScosm}
 \begin{eqnarray}
  [f]_{x}&=&f(x+0)-f(x-0)\label{eqn12}
 \end{eqnarray}
 for an arbitrary function $f$.
 From Eqs.(\ref{eqn7}), (\ref{eqn8}), and (\ref{eqn10}), 
 the integration over $y\in (-0,+0)$ yields
 \begin{eqnarray}
 -3\frac{[\ap]_{0}}{a_{0}}-D\frac{[\cp]_{0}}{c_{0}}
   &=&\kappa^{2}b_{0}\left(V_{1}+\rho_{1}\right)\,,\nonumber\\
   2\frac{[\ap]_{0}}{a_{0}}+\frac{[\np]_{0}}{n_{0}}
   +D\frac{[\cp]_{0}}{c_{0}}
   &=&-\kappa^{2}b_{0}\left(V_{1}-p_{1}\right)\,,\nonumber\\
   3\frac{[\ap]_{0}}{a_{0}}+\frac{[\np]_{0}}{n_{0}}
   +(D-1)\frac{[\cp]_{0}}{c_{0}}
   &=&-\kappa^{2}b_{0}\left(V^{\ast}_{1}-p^{\ast}_{1}\right)\,.\label{eqn13}
 \end{eqnarray}
 Here we use the notation
 $n_{0}=n(t,0)$, $a_{0}=a(t,0)$, $b_{0}=b(t,0)$, and $c_{0}=c(t,0)$.
 The jump conditions on $n,a,$ and $c$ are rewritten as
 \begin{eqnarray}
 \frac{[\ap]_{0}}{a_{0}}&=&
  -\frac{\kappa^{2}b_{0}}{D+3}
   \left(\frac{}{}V_{1}+\rho_{1}
          -D[V_{1}-V^{\ast}_{1}-p_{1}+p^{\ast}_{1}]\right)\,,\nonumber\\
  \frac{[\np]_{0}}{n_{0}}&=&
  -\frac{\kappa^{2}b_{0}}{D+3}
   \left(\frac{}{}V_{1}-2\rho_{1}-3p_{1}
          -D[V_{1}-V^{\ast}_{1}+\rho_{1}+p^{\ast}_{1}]\right)
   \,,\nonumber\\
  \frac{[\cp]_{0}}{c_{0}}&=&
  -\frac{\kappa^{2}b_{0}}{D+3}
   \left(\frac{}{}
        4V_{1}-3V^{\ast}_{1}+\rho_{1}-3p_{1}+3p^{\ast}_{1}
        \right)\,.\label{eqn14}
 \end{eqnarray}
 It is noted that the above jump conditions at $y=0$ depend on the 
 tension, the density, and the pressure on the brane as well as
 on the number of extra dimensions.
 Similarly, the jump conditions at $y=L$ can be derived.
 As mentioned later, these jump conditions are used to derive the
 relations between the brane tensions.  

 In the Randall-Sundrum model, it is important to study the equation 
 for the cosmological expansion on the brane \cite{RScosm}.  
 Below, we consider the simplest case where two branes are completely
 isolated from each other and give the equations of the higher dimensional
 cosmological evolution on the brane.
 This situation corresponds to the limit of $L\rightarrow \infty$.
 Using the jump conditions in Eq.(\ref{eqn14}), the difference 
 between $y=+0$ and $y=-0$ in Eq.(\ref{eqn11}) leads to
 energy conservation
 \begin{eqnarray}
 &&\dot{\rho}_{1}+\frac{9}{D+3}
  \left(\rho_{1}+p_{1}
  +\frac{D}{9}\left[4V_{1}-3V^{\ast}_{1}+4\rho_{1}+3p^{\ast}_{1}\right]
  \right)\frac{\ad_{0}}{a_{0}}\nonumber\\
  &&-D\left\{\frac{}{}
       V_{1}-2\rho_{1}-3p_{1}
        -D\left[
           V_{1}-V^{\ast}_{1}+\rho_{1}+p^{\ast}_{1}
          \right]
       \right\}
     \frac{\cd_{0}}{c_{0}}=0\,\label{eqn15}
 \end{eqnarray}
 on the brane at $y=0$.
 In the case $D=0$, the above equation is reduced to the 
 energy conservation of the four-dimensional standard cosmology.
 We now define the average function \cite{RScosm}
 \begin{eqnarray}
  \{ f \}_{x}&=&\frac{f(x+0)+f(x-0)}{2}\,.\label{eqn16}
 \end{eqnarray}
 Taking the average between $y=+0$ and $y=-0$ 
 with respect to the $(4,4)$ component,
 we can obtain the Friedmann-type equation on the brane at $y=0$:
 \begin{eqnarray}
&&
  \frac{1}{n^{2}_{0}}
  \left[\;
        \frac{\add_{0}}{a_{0}}+\left(\frac{\ad_{0}}{a_{0}}\right)^{2}
       +\frac{D}{3}\frac{\cdd_{0}}{c_{0}}
       +\frac{1}{6}D(D-1)\left(\frac{\cd_{0}}{c_{0}}\right)^{2}
       -\frac{\nd_{0}}{n_{0}}\frac{\ad_{0}}{a_{0}}
       -\frac{D}{3}\frac{\nd_{0}}{n_{0}}\frac{\cd_{0}}{c_{0}}
       +D\frac{\ad_{0}}{a_{0}}\frac{\cd_{0}}{c_{0}}
  \;\right]\nonumber\\
 &&=\frac{1}{3}\kappa^{2}\Lambda\nonumber\\
 &&
 +\frac{\kappa^{4}}{4(D+3)^{2}}
  \left[\frac{}{}
        V_{1}+\rho_{1}-D(V_{1}-V^{\ast}_{1}-p_{1}+p^{\ast}_{1})
  \right]\nonumber\\
 &&\hspace{3cm}\times
  \left[\frac{}{}
        2V_{1}-\rho_{1}-3p_{1}+D(2V_{1}-V^{\ast}_{1}-2p_{1}+p^{\ast}_{1})
  \right]\nonumber\\
 && 
 +\frac{D\kappa^{4}}{24(D+3)^{2}}
   \left[\frac{}{}
        4V_{1}-3V^{\ast}_{1}+\rho_{1}-3p_{1}+3p^{\ast}_{1}
  \right]\nonumber\\
 &&\hspace{3cm}\times
  \left[
        -2V_{1}+3V^{\ast}_{1}-5\rho_{1}-3p_{1}-3p^{\ast}_{1}
        +D(2V_{1}-V^{\ast}_{1}-\rho_{1}-3p_{1}+p^{\ast}_{1})\frac{}{}
  \right]\nonumber\\
 &&
 +\left(1-3\frac{V_{1}-p_{1}}{V_{1}+\rho_{1}}\right)
  \left(\frac{\{\ap\}_{0}}{a_{0}b_{0}}\right)^{2}
 +\frac{D}{6}
  \left(D-1-2D\frac{V^{\ast}_{1}-p_{1}}{V^{\ast}_{1}+\rho_{1}}\right)
  \left(\frac{\{\cp\}_{0}}{c_{0}b_{0}}\right)^{2}\nonumber\\
 &&
 +D\left(1-\frac{V_{1}+V^{\ast}_{1}-p_{1}-p^{\ast}_{1}}
                          {V_{1}+\rho_{1}}\right)
  \frac{\{\cp\}_{0}\{\ap\}_{0}}{a_{0}c_{0}b^{2}_{0}}\,.\label{eqn17}
 \end{eqnarray}
 Imposing the orbifold symmetry $y\sim -y$, we have $\{f^{\prime}\}=0$.
 Then we can drop all terms involving the average in Eq.(\ref{eqn17}).
 We have also fixed the time in such a way that $n_{0}=1$. This 
 corresponds to the usual choice of time in conventional cosmology.
 We introduce the Hubble parameters $H_{a}\equiv \dot{a}/a$
 and $H_{c}\equiv \dot{c}/c$ for the two scale factors on the brane
 at $y=0$.
 Here it is assumed that after stabilizing the radion $b_{0}$
 the matter on the brane is isotropic and that radiation dominates.
 This leads to $p_{1}=p^{\ast}_{1}$ and $\rho_{1}=(D+3)p_{1}$.
 Then the cosmological evolution equation on the brane at $y=0$ becomes
 \begin{eqnarray}
 &&\dot{H_{a}}+2H_{a}^{2}
       +\frac{D}{3}\dot{H_{c}}
       +\frac{1}{6}D(D+1)H_{c}^{2}
       +DH_{a}H_{c}\nonumber\\
 &&=\frac{1}{3}\kappa^{2}\L
    +\frac{\kappa^{4}}{4(D+3)^{2}}
  \left[\frac{}{}
        V_{1}+\rho_{1}-D(V_{1}-V^{\ast}_{1})
  \right]\nonumber \\
 &&\hspace{3cm}
 \times
  \left[
        2V_{1}-\rho_{1}\frac{6+D}{3+D}
   +D\left(2V_{1}-V^{\ast}_{1}-\rho_{1}\frac{1}{3+D}\right)
  \right]\nonumber\\
 && 
 +\frac{D\kappa^{4}}{24(D+3)^{2}}
   \left[\frac{}{}
        4V_{1}-3V^{\ast}_{1}+\rho_{1}
  \right]\nonumber\\
 &&\hspace{2cm}\times
  \left[
        -2V_{1}+3V^{\ast}_{1}-\rho_{1}\frac{21+D}{3+D}
        +D\left(2V_{1}-V^{\ast}_{1}-\rho_{1}\frac{5+D}{3+D}\right)
  \right]\,.
    \label{eqn18} 
 \end{eqnarray}
 The energy conservation equation and the cosmological evolution equation
 on the brane at $y=L$ can be obtained by the same procedures 
 as mentioned above.

 Since the cosmology equation obtained here corresponds to a
 completely isolated brane system, 
 it implies that the matter on one brane has nothing to do with 
 the matter on another brane.
 However, in the case of finite $L$, the two branes are closely related, 
 so that
 the matter on one brane is constrained by the matter on the other brane.
 To study the cosmology equations constrained by two branes \cite{RScosm1},
 the relation between the functions $n$, $a$, $b$, and $c$ must obtained by 
 integrating out Eq.(\ref{eqn11}).
 As for this point, we are going to provide the analysis in detail of
 the cosmological evolution in the setup presented in this paper
 \cite{RScosm2}. 
%
%%%%%%%%%%%%%%%%%%%%%%%%%%% 3 %%%%%%%%%%%%%%%%%%%%%%%%%%%%%%%%%%%%
 \section{Static solutions}
%%%%%%%%%%%%%%%%%%%%%%%%%%%%%%%%%%%%%%%%%%%%%%%%%%%%%%%%%%%%%%%%%%
 $\hspace{0.5cm}$
 We can obtain the static Randall-Sundrum-type solution by setting
 the density and the pressure of matter to zero.
 Note that the functions $n,a,$ and $c$ have time independence and preserve
 Poincar$\acute{\rm e}$ invariance in the $(1+3)$-dimensional metric
 \begin{eqnarray}
  n(y)=a(y),\hspace{1cm}b=1\,,\label{eqn19} 
 \end{eqnarray}
 where $b$ is normalized to be unity since it is assumed that
 the size in the $y$ direction compactified on the orbifold is
 stabilized via some mechanism.
 We consider that the four-dimensional warped metric function $a(y)$ is
 generally  different from the extra $D$-dimensional $c(y)$.
 Following from Eq.(\ref{eqn7}) to Eq.(\ref{eqn10}), 
 the Einstein equations in the bulk are given by
 \begin{eqnarray}
  &&12\left(\frac{\ap}{a}\right)^{2}
 +D(D-1)\left(\frac{\cp}{c}\right)^{2}
 +8D\frac{\ap}{a}\frac{\cp}{c}
 =-2\kappa^{2}\Lambda\,,\label{eqn20} \\
 &&6\left(\frac{\ap}{a}\right)^{2}
 +D(D-1)\left(\frac{\cp}{c}\right)^{2}
 +6\frac{\app}{a}+2D\frac{\cpp}{c}
 +6D\frac{\ap}{a}\frac{\cp}{c}
 =-2\kappa^{2}\Lambda\,,\label{eqn21} \\
 &&12\left(\frac{\ap}{a}\right)^{2}
 +(D-1)(D-2)\left(\frac{\cp}{c}\right)^{2}
 +8\frac{\app}{a}+2(D-1)\frac{\cpp}{c}
 +8(D-1)\frac{\ap}{a}\frac{\cp}{c}
 =-2\kappa^{2}\Lambda\,,\nonumber\\\label{eqn22} 
 \end{eqnarray}
 where we used the fact that the $(0,0)$ component is equal to
 the $(a,a)$ component.

 Here we can derive the solution of the five-dimensional classical 
 Einstein equation to be the Randall-Sundrum model.
 Setting $D=0$ and neglecting Eq.(\ref{eqn22}) coming from
 the $(a,a)$ component for the extra $D$ dimensions, we have
 \begin{eqnarray}
  \left(\frac{\ap}{a}\right)^{2}=
  \frac{\app}{a}=-\frac{\k^{2}\L}{6}\,.\label{eqn23}
 \end{eqnarray}
 For $\L<0$, the $S^{1}/Z_{2}$ orbifold symmetric solution is of the form
 \begin{eqnarray}
  a(y)=e^{\pm m_{0}|y|}\,,\label{eqn24}
 \end{eqnarray}
 where
 \begin{eqnarray}
  m_{0}=\sqrt{\frac{-\k^{2}\L}{6}}\,.\label{eqn25}
 \end{eqnarray}
 From Eq.(\ref{eqn14}), the jump conditions of $a(y)$ at $y=0$ and $y=L$
 lead to
 \begin{eqnarray}
  V_{1}=-V_{2}=\mp \sqrt{\frac{-\L}{6\k^{2}}}\,,\label{eqn26}
 \end{eqnarray}
 where upper and lower signs correspond to the signs
 in Eq.(\ref{eqn24}), respectively.  
 Thus the brane tensions $V_{1}$, $V_{2}$ at 
 $y=0$ and $y=L$ have opposite sign from each other.
 When $V_{2}$ is negative, the warped metric becomes
 \begin{eqnarray}
  ds^{2}=e^{- 2m_{0}|y|}g_{\mu\nu}dx^{\mu}dx^{\nu}-dy^{2}
  \label{eqn27}\,,
 \end{eqnarray}
 where $g_{\mu\nu}={\rm diag}(+,-,-,-)$.
 By using this warped metric, 
 Randall and Sundrum proposed an alternative solution to the hierarchy 
 problem.
 This solution appeals to the possibility of an extra dimension
 limited by two branes with positive and negative tensions.
 Further, the resolution of the hierarchy problem is possible 
 provided that the observable brane at $y=L$ is the one with
 the negative tension.
 This model insists that the hierarchy has its origin in
 the geometry of the extra dimension.

 We are interested in the solutions of the Randall-Sundrum model 
 embedded in $5+D$ dimensions with an orbifold compactification.
 The feature of this setup is that the warped metric function $a(y)$ of 
 the four-dimensional spacetime
 is generally different from the one $c(y)$ of the extra $D$-dimensional
 space.
 After some algebra,
 we can rewrite the appropriate linear combination of Eqs.(\ref{eqn20}), 
 (\ref{eqn21}), and (\ref{eqn22}) as
 \begin{eqnarray}
  12\left(\frac{\ap}{a}\right)^{2}
 +D(D-1)\left(\frac{\cp}{c}\right)^{2}
 +8D\frac{\ap}{a}\frac{\cp}{c}&=& -2\k^{2}
   \Lambda\,,\label{eqn28} \\
  \frac{\app}{a}+3\left(\frac{\ap}{a}\right)^{2}
  +D\frac{\ap}{a}\frac{\cp}{c}&=& -\frac{2\k^{2}}{D+3}
  \Lambda\,,\label{eqn29} \\
  \frac{\cpp}{c}+(D-1)\left(\frac{\cp}{c}\right)^{2}
  +4\frac{\ap}{a}\frac{\cp}{c}&=& -\frac{2\k^{2}}{D+3}
    \Lambda\,.\label{eqn30} 
 \end{eqnarray}

 To study the behavior of $a(y)$ and $c(y)$, let us consider 
 the three cases of $\L<0$, $\L>0$, and $\L=0$ separately.
%%%%%%%%%%%%%%%%%%%%%%%%%%% 3.1 %%%%%%%%%%%%%%%%%%%%%%%%%%%%%%%%%
 \subsection{The solutions for $\L<0$}
%%%%%%%%%%%%%%%%%%%%%%%%%%%%%%%%%%%%%%%%%%%%%%%%%%%%%%%%%%%%%%%%%
%  
 $\hspace{0.5cm}$
 In this case, in order to obtain the solution of the Einstein equations,
 we perform the changes of variables that allow the exact solution
 Eqs.(\ref{eqn28})-(\ref{eqn30}). 
 We define $A(y)$ and $C(y)$ by
 \begin{eqnarray}
 a(y)=e^{A(y)},\hspace{0.5cm}c(y)=e^{C(y)}\,.\label{eqn31}
 \end{eqnarray}
 Further, defining the parameter $\omega\equiv 2\k^{2}$,
 from Eqs.(\ref{eqn28}), (\ref{eqn29}) and (\ref{eqn30}), we obtain
 \begin{eqnarray}
  12(\Ap)^{2}+D(D-1)(\Cp)^{2}+8D\Ap\Cp&=&-\omega\L\nonumber\,,\\
  \App+4(\Ap)^{2}+D\Ap\Cp&=&-\frac{\omega\L}{D+3}\nonumber\,,\\
  \Cpp+D(\Cp)^{2}+4\Ap\Cp&=&-\frac{\omega\L}{D+3}\,.\label{eqn32}
 \end{eqnarray}
 We introduce the new variable $Y$ 
 \begin{eqnarray}
  dY=\mp e^{-4A-DC}dy\label{eqn33}\,.
 \end{eqnarray}
 This replacement of the variable is similar to the procedure of 
 seeking the Kasner solution with time dependence in the 
 higher dimensional cosmology in Ref.\cite{LEcosm}.
 Thus we get 
 \begin{eqnarray}
  12\left(\frac{dA}{dY}\right)^{2}
 +D(D-1)\left(\frac{dC}{dY}\right)^{2}
 +8D\frac{dA}{dY}\frac{dC}{dY}&=&-\omega\L e^{8A+2DC}\,,
  \label{eqn34}\\
 \frac{d^{2}A}{dY^{2}}=\frac{d^{2}C}{dY^{2}}&=&
 -\frac{\omega\L}{D+3}e^{8A+2DC}\,.\label{eqn35}
 \end{eqnarray}
 Note that the above equations are unchanged in either case
 when the lower or upper sign  included in Eq.(\ref{eqn33}) is taken.
 Hence, we can immediately write the integral of Eq.(\ref{eqn35}):
 \begin{eqnarray}
  A-C=P_{1}Y+P_{2}\label{eqn36}\,,
 \end{eqnarray}
 where $P_{1}$ and $P_{2}$ are the integration constants.
 To solve the differential equation of Eq.(\ref{eqn34}),
 we define a new variable \cite{LEcosm}
 \begin{eqnarray}
  Z=8A+2DC\,,\label{eqn37}
 \end{eqnarray}
 Equation (\ref{eqn34}) is translated as
 \begin{eqnarray}
  \frac{d^{2}Z}{dY^{2}}&=&-\frac{2(D+4)\omega\L}{D+3}e^{Z}\label{eqn38}\,.
 \end{eqnarray}
 Integrating this equation, we obtain
 \begin{eqnarray}
  \left(\frac{dZ}{dY}\right)^{2}=
  -\frac{4(D+4)\omega\L}{D+3}e^{Z}+P_{3}\,,\label{eqn39}
 \end{eqnarray}
 where $P_{3}$ is the integration constant.
 However, $P_{3}$ is a function of $P_{1}$.
 This is because substitution of Eq.(\ref{eqn37}) into Eq.(\ref{eqn39})
 leads to
 \begin{eqnarray}
  P_{3}=\frac{16D}{D+3}\left(\frac{dA}{dY}-\frac{dC}{dY}\right)^{2}
       =\frac{16D}{D+3}P^{2}_{1}\,.\label{eqn40}
 \end{eqnarray}
 Therefore, we have
 \begin{eqnarray}
  \left(\frac{dZ}{dY}\right)^{2}=
  -\frac{4(D+4)\omega\L}{D+3}e^{Z}+\frac{16D}{D+3}P^{2}_{1}\,.
  \label{eqn41}
 \end{eqnarray}
 Since $P_{1}$ and $P_{2}$ are determined by the initial condition,
 the values are expected to be determined via some dynamics of
 the underlying physics. 
 Below, we point out that different types of 
 solution of Eq.(\ref{eqn41}) are obtained depending on
 whether $P_{1}$ is nonvanishing or vanishing.

 First, we consider the case of $P_{1}\neq 0$ for the negative
 bulk cosmological constant.
 Equation (\ref{eqn41}) can be simply solved as
 \begin{eqnarray}
  e^{Z}=\frac{4DP^{2}_{1}}{(D+4)\omega |\L|}
          \frac{1}{\sinh^{2}\left(
          2\sqrt{\frac{D}{D+3}}P_{1}Y\right)}\label{eqn42}\,.
 \end{eqnarray}
 Using Eqs.(\ref{eqn36}) and (\ref{eqn37}), we get
 \begin{eqnarray}
  a=e^{A}&=&\left[
          \frac{4DP^{2}_{1}}{(D+4)\omega |\L|}
          \frac{e^{2D(P_{1}Y+P_{2})}}{\sinh^{2}\left(
          2\sqrt{\frac{D}{D+3}}P_{1}Y\right)}
          \right]^{\frac{1}{2(4+D)}}\,,\nonumber\\
  && \nonumber\\
  c=e^{C}&=&\left[
          \frac{4DP^{2}_{1}}{(D+4)\omega |\L|}
          \frac{e^{-8(P_{1}Y+P_{2})}}{\sinh^{2}\left(
          2\sqrt{\frac{D}{D+3}}P_{1}Y\right)}
          \right]^{\frac{1}{2(4+D)}}\label{eqn43}\,.
 \end{eqnarray}
 To write the above equations in terms of $y$, we need to change
 the variable $Y$ into $y$. Equation (\ref{eqn33}) becomes
 \begin{eqnarray}
  dY=\mp\sqrt{\frac{D+4}{4DP^{2}_{1}}\omega|\L|}\;
     \sinh\left(2\sqrt{\frac{D}{D+3}}\;P_{1}Y\right)\;dy\,
     \label{eqn44}
 \end{eqnarray}
 which leads to the relation between $Y$ and $y$: 
 \begin{eqnarray}
 \e\left[\;
     2\sqrt{\frac{D}{D+3}}P_{1}Y
     \right]
 =
  \coth\left(\frac{1}{2}\sqrt{\frac{D+4}{D+3}\omega|\L|}\;(y+y_{0})\right)
  \,,\label{eqn45}
 \end{eqnarray}
 where $y_{0}$ is an integration constant and the positive sign in 
 Eq.(\ref{eqn44}) is taken.
 After substitution of Eq.(\ref{eqn45}) into Eq.(\ref{eqn43}),
 the functions $a$ and $c$ are described in terms of $y$:
 \begin{eqnarray}
  a(y)&=&
      \left(\frac{4DP^{2}_{1}}{(D+4)\omega|\L|}
      \right)^{\frac{1}{2(D+4)}}\nonumber\\
  &&\times
      \left[\;
      \coth^{\frac{\sqrt{D(D+3)}}{2}}
      \left(
       \frac{1}{2}\sqrt{\frac{D+4}{D+3}\omega|\L|}\;(|y|+y_{0})
      \right)
      \sinh\left(\sqrt{\frac{D+4}{D+3}\omega|\L|}\;(|y|+y_{0})\right)
      \;\right]^{\frac{1}{D+4}}\,,\nonumber\\
  c(y)&=&
      \left(\frac{4DP^{2}_{1}}{(D+4)\omega|\L|}
      \right)^{\frac{1}{2(D+4)}}\nonumber\\
  &&\times
      \left[\;
      \tanh^{2\sqrt{\frac{D+3}{D}}}
      \left(
       \frac{1}{2}\sqrt{\frac{D+4}{D+3}\omega|\L|}\;(|y|+y_{0})
      \right)
      \sinh\left(\sqrt{\frac{D+4}{D+3}\omega|\L|}\;(|y|+y_{0})\right)
      \;\right]^{\frac{1}{D+4}}\,,\nonumber\\\label{eqn46}
 \end{eqnarray}
 where $a$ and $c$ respect the $Z_{2}$ symmetry $y\sim -y$.
 The constant $P_{2}$ in Eq.(\ref{eqn36}) does not appear in the expression
 for the metric functions $a(y)$ and $c(y)$, namely, $P_{2}$ could be
 set to zero
 since it can be absorbed into a redefinition of the extra 
 $D$-dimensional coordinates.
 Moreover, the coefficient without $P_{2}$
 in $a(y)$ and $c(y)$ is considered
 to be physical irrelevant since we are interested only in the 
 behavior with respect to $y$.
 From Eq.(\ref{eqn14}),  
 the jump conditions at $y=0$ lead to
 \begin{eqnarray}
  \frac{2\sqrt{|\l|}}{4+D}
  \frac{2\cosh\sqrt{|\l|} y_{0}-\sqrt{D(D+3)}}{2\sinh\sqrt{|\l|} y_{0}}
 &=&-\frac{\kappa^{2}}{D+3}
                      \left(V_{1}-D[V_{1}-V^{\ast}_{1}]\right)
 \,,\nonumber\\
 \frac{2\sqrt{|\l|}}{4+D}
  \frac{\cosh\sqrt{|\l|} y_{0}+2\sqrt{\frac{D+3}{D}}}
  {\sinh\sqrt{|\l|} y_{0}}
 &=&-\frac{\kappa^{2}}{D+3}
                      \left(4V_{1}-3V^{\ast}_{1}\right)\,,\label{eqn47}
 \end{eqnarray}
 where we define
 \begin{eqnarray}
  \l=\frac{D+4}{D+3}\omega\L\,.\label{eqn48}
 \end{eqnarray}
 Furthermore, the jump conditions at $y=L$ lead to
 \begin{eqnarray}
  \frac{2\sqrt{|\l|}}{4+D}
  \frac{\sinh\sqrt{|\l|} y_{0}\;(2\cosh\sqrt{|\l|} y_{0}-\sqrt{D(D+3)}
  \cosh\sqrt{|\l|} L)}
       {\cosh 2\sqrt{|\l|} y_{0}-\cosh 2\sqrt{|\l|} L}
  &=&\frac{\kappa^{2}}{D+3}
     \left(V_{2}-D[V_{2}-V^{\ast}_{2}]\right)\,,\nonumber\\
  \frac{2\sqrt{|\l|}}{4+D}
  \frac{2\sinh\sqrt{|\l|} y_{0}\;(\cosh\sqrt{|\l|} y_{0}
  +2\sqrt{\frac{D+3}{3}}\cosh\sqrt{|\l|} L)}
       {\cosh 2\sqrt{|\l|} y_{0}-\cosh 2\sqrt{|\l|} L}
  &=&\frac{\kappa^{2}}{D+3}
          (4V_{2}-3V^{\ast}_{2})\,.\label{eqn49}
 \end{eqnarray}
 From Eqs.(\ref{eqn47}) and (\ref{eqn49}), 
 the brane tensions at $y=0$ and $y=L$ are expressed as
 \begin{eqnarray}
  V_{1}&=&-\frac{2\sqrt{|\l|}}{\k^{2}(4+D)}
           \frac{2(D+3)\cosh\sqrt{|\l|} y_{0}+\sqrt{D(D+3)}}
                 {2\sinh\sqrt{|\l|} y_{0}}\,,\nonumber\\
  V^{\ast}_{1}&=&-\frac{2\sqrt{|\l|}}{\k^{2}(4+D)}
                  \frac{(D+3)\cosh\sqrt{|\l|} y_{0}-2\sqrt{\frac{D+3}{D}}}
                        {\sinh\sqrt{|\l|} y_{0}}\,,\nonumber\\
  V_{2}&=& \frac{2\sqrt{|\l|}}{\k^{2}(4+D)}
  \frac{\{2(D+3)\cosh\sqrt{|\l|} y_{0}+\sqrt{D(D+3)}\cosh\sqrt{|\l|} L\}
  \sinh\sqrt{|\l|} y_{0}}
     {\cosh 2\sqrt{|\l|} y_{0}-\cosh 2\sqrt{|\l|} L}\,,\nonumber\\
  V^{\ast}_{2}&=&\frac{2\sqrt{|\l|}}{\k^{2}(4+D)}
 \frac{2\{(D+3)\cosh\sqrt{|\l|} y_{0}-2\sqrt{\frac{D+3}{D}}\cosh\sqrt{|\l|} L\}
 \sinh\sqrt{|\l|} y_{0}}
      {\cosh 2\sqrt{|\l|} y_{0}-\cosh 2\sqrt{|\l|} L}\,.\label{eqn50}
 \end{eqnarray}
 From the above equations, the integration constant $y_{0}$ is expressed as
 \begin{eqnarray}
 y_{0}=\frac{1}{\sqrt{|\l|}}\;
 {\rm arc}\sinh\left[\;
         \frac{1}{\k(V^{\ast}_{1}-V_{1})}
         \sqrt{\frac{2(D+4)}{D}|\L|}
         \;\right]\,.
  \label{eqn51}
 \end{eqnarray} 
 The sign of $y_{0}$ depends on the sign of the difference 
 between $V_{1}$ and $V^{\ast}_{1}$.
 If $V_{1}>V^{\ast}_{1}$, $y_{0}$ becomes negative,
 and Eq.(\ref{eqn46}) leads to the conclusion that
 $a(y)$ has a singular point as long as $D\neq 1$.
 To avoid this a singular point over $|y|\leq L$, $-y_{0}>L$ is required.
 
 From Eq.(\ref{eqn50}), the brane tensions $V_{2}$,
 $V^{\ast}_{2}$ of the observable brane at $y=L$
 can be described in terms of $V_{1}$, $V^{\ast}_{1}$ of the hidden 
 brane at $y=0$.
 The ratios $V_{1}/V^{\ast}_{1}$ and $V_{2}/V^{\ast}_{2}$ cannot be
 unity as long as $D$ is a positive integer.
 Thus, each brane tension
 becomes anisotropic in this setup, and
 each brane tension is closely related to the other
 because of the presence of two branes.
 Taking the limit $L\gg y_{0}$ in the infinite orbifold 
 extra dimension
 where the observable brane is fixed far away from the origin 
 \cite{infRS},
 the ratio $V_{2}$ to $V^{\ast}_{2}$ of the observable brane becomes
 \begin{eqnarray}
  \frac{V_{2}}{V^{\ast}_{2}}=-\frac{1}{4}D\,.\label{eqn52}
 \end{eqnarray} 
 $V_{2}$ and $V^{\ast}_{2}$ have opposite signs to each other and 
 the magnitude of the ratio depends on the number of extra $D$ dimensions.

 Next, let us consider the case $P_{1}= 0$.
 There exists a solution of Eq.(\ref{eqn41}) with a negative 
 bulk cosmological constant. Integrating it we obtain
 \begin{eqnarray}
  e^{Z}=\frac{D+3}{(D+4)\omega |\L|Y^{2}}\,.\label{eqn53}
 \end{eqnarray}
 Using Eqs.(\ref{eqn36}) and (\ref{eqn37}), we have
 \begin{eqnarray}
  e^{A}&=&
  \left[
          \frac{D+3}{(D+4)\omega |\L|}
          \frac{e^{2DP_{2}}}{Y^{2}}
          \right]^{\frac{1}{2(4+D)}}\,,\nonumber\\
  && \nonumber\\
  e^{C}&=&\left[
          \frac{D+3}{(D+4)\omega |\L|}
          \frac{e^{-8P_{2}}}{Y^{2}}
          \right]^{\frac{1}{2(4+D)}}\,.\label{eqn54}
 \end{eqnarray}
 Equation (\ref{eqn33}) leads to the relation between $Y$ and $y$: 
 \begin{eqnarray}
  Y=Y_{0}\;\e\left[
      \mp\sqrt{\frac{D+4}{D+3}\omega |\L|}\;y
      \right]\,,\label{eqn55}
 \end{eqnarray}
 where $Y_{0}$ is an integration constant.
 After substitution of Eq.(\ref{eqn55}) into Eq.(\ref{eqn54}),
 $a$ and $c$ are described in terms of $y$:
 \begin{eqnarray}
  a(y)=c(y)=
          \e\left[
            \pm\sqrt{\frac{\omega |\L|}{(D+3)(D+4)}}\;|y|
            \right]
          \,.\label{eqn56}
 \end{eqnarray}
 Hence the upper and lower signs correspond to the signs in 
 Eq.(\ref{eqn55}) and both $a$ and $c$ are normalized to be unity 
 at $y=0$.
 The jump conditions of $y=0$ and $y=L$ lead to
 \begin{eqnarray}
  \pm 2\sqrt{\frac{\omega |\L|}{(D+3)(D+4)}}&=&
  -\frac{\kappa^{2}}{D+3}
                      \left(V_{1}-D[V_{1}-V^{\ast}_{1}]\right)=
  -\frac{\kappa^{2}}{D+3}
                      \left(4V_{1}-3V^{\ast}_{1}\right)\nonumber\\
 &=&\frac{\kappa^{2}}{D+3}
                      \left(V_{2}-D[V_{2}-V^{\ast}_{2}]\right)=
  \frac{\kappa^{2}}{D+3}
                      \left(4V_{2}-3V^{\ast}_{2}\right)\,,\label{eqn57}  
 \end{eqnarray}
 then the brane tensions are given by
 \begin{eqnarray}
  V_{1}=V^{\ast}_{1}=-V_{2}=-V^{\ast}_{2}=
  \mp \frac{2}{\k}\sqrt{2\frac{D+3}{D+4}|\L|}\,.\label{eqn58}
 \end{eqnarray}
 In this case, we find that $a(y)$ is equal to $c(y)$ and
 the brane tension of each brane 
 is automatically guaranteed to be isotopic.
 The lower sign in Eq.(\ref{eqn58}) corresponds to the case
 that the brane tension of the observable brane at $y=L$
 is negative.
 Consequently, the warped metric function becomes the exponential damping 
 factor since the lower sign (negative) in Eq.(\ref{eqn56}) is selected.
 This situation is similar to the five-dimensional Randall-Sundrum 
 solution.
 As in the Randall-Sundrum scenario, the hierarchy between the physical mass
 scale $m_{\rm hid}$ on the hidden brane at $y=0$ and $m_{\rm obs}$ 
 on the observable
 brane at $y=L$ can be generated from the warped metric.

 In the case of a negative bulk cosmological constant, 
 whether the integration constant $P_{1}$ is nonzero or zero determines
 the form of the warped metric function.
 If $P_{1}\neq 0$, the warped metric functions $a(y)$ and $c(y)$ have
 the forms of different hyperbolic functions and the brane
 tension is anisotropic.
 If $P_{1}=0$ and the observable brane has negative brane tension,
 both the warped metric functions have the same form of 
 exponential damping factor and the brane tension is isotropic.
 Thus, whether the brane tension on the orbifold is isotropic
 or anisotropic depends on the value of $P_{1}$,
 namely, the integration constant $P_{1}$ controls the solution of the
 Einstein equation in bulk and it is expected that $P_{1}$ is determined by 
 the initial configuration of the brane world in this setup.
 Hence it is assumed that the dynamical mechanism for fixing the 
 value of $P_{1}$ is unknown.
%    
%%%%%%%%%%%%%%%%%%%%%%%%%%% 3.3 %%%%%%%%%%%%%%%%%%%%%%%%%%%%%%%%%
 \subsection{The solutions for $\L>0$}
%%%%%%%%%%%%%%%%%%%%%%%%%%%%%%%%%%%%%%%%%%%%%%%%%%%%%%%%%%%%%%%%%
% 

 When the bulk cosmological constant $\L$ is positive, 
 Eq.(\ref{eqn41}) implies that $P_{1}$ should be nonzero.
 Solving this equation, we obtain
 \begin{eqnarray}
  e^{Z}=\frac{4DP^{2}_{1}}{(D+4)\omega \L}
          \frac{1}{\cosh^{2}\left(
          2\sqrt{\frac{D}{D+3}}P_{1}Y\right)}\label{eqn59}\,.
 \end{eqnarray} 
 Furthermore, we have
 \begin{eqnarray}
  e^{A}&=&\left[
          \frac{4DP^{2}_{1}}{(D+4)\omega \L}
          \frac{e^{2D(P_{1}Y+P_{2})}}{\cosh^{2}\left(
          2\sqrt{\frac{D}{D+3}}P_{1}Y\right)}
          \right]^{\frac{1}{2(4+D)}}\,,\nonumber\\
  && \nonumber\\
  e^{C}&=&\left[
          \frac{4DP^{2}_{1}}{(D+4)\omega \L}
          \frac{e^{-8(P_{1}Y+P_{2})}}{\cosh^{2}\left(
          2\sqrt{\frac{D}{D+3}}P_{1}Y\right)}
          \right]^{\frac{1}{2(4+D)}}\label{eqn60}\,.
 \end{eqnarray}
 Following Eq.(\ref{eqn33}),
 \begin{eqnarray}
 \e\left[
   2\sqrt{\frac{D}{D+3}}P_{1}Y
   \right]=
 \mp\tan\left(\frac{1}{2}
            \sqrt{\frac{D+4}{D+3}\omega\L}\;(y+y_{1})\right)\,,\label{eqn61}
 \end{eqnarray}
 where $y_{1}$ is an integration constant.
 By imposing the orbifold symmetry, we can rewrite $a$ and $c$ in terms of $y$
 \begin{eqnarray}
  a(y)&=&
      \left(\frac{4DP^{2}_{1}}{(D+4)\omega\L}
      \right)^{\frac{1}{2(D+4)}}
      \left[\;
      \tan^{\frac{\sqrt{D(D+3)}}{2}}
      \left(
       \frac{\sqrt{\l}}{2}\;(|y|+y_{1})
      \right)\;
      \sin\left(\sqrt{\l}\;(|y|+y_{1})\right)
      \;\right]^{\frac{1}{D+4}}\,,\nonumber\\
  c(y)&=&
      \left(\frac{4DP^{2}_{1}}{(D+4)\omega\L}
      \right)^{\frac{1}{2(D+4)}}
      \left[\;
      \cot^{2\sqrt{\frac{D+3}{D}}}
      \left(
       \frac{\sqrt{\l}}{2}\;(|y|+y_{1})
      \right)\;
      \sin\left(\sqrt{\l}\;(|y|+y_{1})\right)
      \;\right]^{\frac{1}{D+4}}\,,\nonumber\\\label{eqn62}
 \end{eqnarray}
 where $\l$ is defined in Eq.(\ref{eqn48}) and $P_{2}$ can be set to
 zero as mentioned previously.
 The jump conditions lead to brane tensions as follows:
 \begin{eqnarray}
  V_{1}&=&-\frac{2\sqrt{\l}}{\k^{2}(4+D)}
           \frac{2(D+3)\cos\sqrt{\l} y_{1}-\sqrt{D(D+3)}}
                 {2\sin\sqrt{\l} y_{1}}\,,\nonumber\\
  V^{\ast}_{1}&=&-\frac{2\sqrt{\l}}{\k^{2}(4+D)}
                  \frac{(D+3)\cos\sqrt{\l} y_{1}+2\sqrt{\frac{D+3}{D}}}
                        {\sin\sqrt{\l} y_{1}}\,,\nonumber\\
  V_{2}&=& \frac{2\sqrt{\l}}{\k^{2}(4+D)}
           \frac{\sinh\sqrt{\l} y_{1}\;
     (2(D+3)\cosh\sqrt{\l} y_{1}+\sqrt{D(D+3)}\cosh\l L)}
                 {\cosh 2\sqrt{\l} y_{1}-\cosh 2\sqrt{\l} L}\,,\nonumber\\
  V^{\ast}_{2}&=&\frac{2\sqrt{\l}}{\k^{2}(4+D)}
 \frac{2\sinh\sqrt{\l} y_{1}\;
  ((D+3)\cosh\sqrt{\l} y_{1}-2\sqrt{\frac{D+3}{D}}\cosh \sqrt{\l}L)}
      {\cosh 2\sqrt{\l} y_{1}-\cosh 2\sqrt{\l} L}\,.\label{eqn63}
 \end{eqnarray}
 Using the above equations, the integration constant $y_{1}$ is expressed as
 \begin{eqnarray}
 y_{1}=\frac{1}{\sqrt{\l}}\;
 \arcsin\left[\;
         \frac{1}{\k(V_{1}-V^{\ast}_{1})}
         \sqrt{\frac{2(D+4)}{D}\L}
         \;\right]\,.
  \label{eqn64}
 \end{eqnarray} 
 The sign of $y_{1}$ depends on the sign of the difference between $V_{1}$
 and $V^{\ast}_{1}$.
 If $V^{\ast}_{1}>V_{1}$, $y_{1}$ becomes negative and it is found that
 $c(y)$ has a singularity.
 Moreover, Eq.(\ref{eqn61}) leads to the constraint on
 the location of $y=L$:
 \begin{eqnarray}
  L<\pi\sqrt{\frac{D+3}{(D+4)\omega\L}}-y_{1}\,.\label{eqn65}
 \end{eqnarray}

 As in the case of negative bulk cosmological constant,
 the brane tension becomes anisotropic.
 The warped metric functions obtained here have the forms of 
 different trigonometric functions.
%
%%%%%%%%%%%%%%%%%%%%%%%%%%% 3.3 %%%%%%%%%%%%%%%%%%%%%%%%%%%%%%%%%
 \subsection{The solutions for $\L=0$}
%%%%%%%%%%%%%%%%%%%%%%%%%%%%%%%%%%%%%%%%%%%%%%%%%%%%%%%%%%%%%%%%%
%
 In the case of zero bulk cosmological constant,
 the metric functions are quite similar to the Kasner 
 solution in higher dimensional cosmology given in the appendix.
 We take the power-law form by taking account of the orbifold symmetry,
 \begin{eqnarray}
  a(y)&=&\left(\frac{|y|}{y_{2}}+1\right)^{k}\,,\nonumber\\
  c(y)&=&\left(\frac{|y|}{y_{2}}+1\right)^{l}\,,\label{eqn66}
 \end{eqnarray}
 where $a$ and $c$ are normalized to be unity at $y=0$ and
 $y_{2}$ is a constant to be determined later.
 Substituting the above equations into Eqs.(\ref{eqn20})-(\ref{eqn22}),
 we obtain two equations for the exponents
 \begin{eqnarray}
  4k+Dl&=&1\,,\nonumber\\
  4k^{2}+Dl^{2}&=&1\,.\label{eqn67}
 \end{eqnarray}  
 Solving the above equations, $k$ and $l$ are given by
 \begin{eqnarray}
  k&=&\frac{2\pm\sqrt{D(D+3)}}{2(D+4)}\,,\nonumber\\
  l&=&\frac{D\mp 2\sqrt{D(D+3)}}{D(D+4)}\,.\label{eqn68}
 \end{eqnarray}
 Here, we cannot determine whether the sign included in Eq.(\ref{eqn68}) 
 is the lower or upper sign at this stage.

 From Eq.(\ref{eqn14}), the jump conditions on $a$ and $c$ at $y=0$ yield
 \begin{eqnarray}
  \frac{2k}{y_{2}}&=&-\frac{\kappa^{2}}{D+3}
                      \left(V_{1}-D[V_{1}-V^{\ast}_{1}]\right)\,,\nonumber\\
  \frac{2l}{y_{2}}&=& -\frac{\kappa^{2}}{D+3}
                      (4V_{1}-3V^{\ast}_{1})\,,
  \label{eqn69}
 \end{eqnarray}
 furthermore, the jump conditions at $y=L$ lead to
 \begin{eqnarray}
  \frac{2ky_{2}}{L^{2}-y^{2}_{2}}&=&-\frac{\kappa^{2}}{D+3}
                      \left(V_{2}-D[V_{2}-V^{\ast}_{2}]\right)\,,\nonumber\\
  \frac{2ly_{2}}{L^{2}-y^{2}_{2}}&=& -\frac{\kappa^{2}}{D+3}
                      (4V_{2}-3V^{\ast}_{2})\,.
  \label{eqn70}
 \end{eqnarray}
 Using the above equations, the brane tensions are given by
 \begin{eqnarray}
  V_{1}&=& -\frac{2}{\k^{2}y_{2}}(1-k)\,,\nonumber\\
  V_{1}^{\ast}&=& -\frac{2}{\k^{2}y_{2}}(1-l)\,,\nonumber\\
  V_{2}&=& -\frac{2y_{2}}{\k^{2}(L^{2}-y^{2}_{2})}(1-k)\,,\nonumber\\
  V_{2}^{\ast}&=& -\frac{2y_{2}}{\k^{2}(L^{2}-y^{2}_{2})}(1-l)\,.
  \label{eqn71}
 \end{eqnarray}
 Following the constraint on the exponents in Eq.(\ref{eqn67}), 
 the constant $y_{2}$ becomes
 \begin{eqnarray}
  y_{2}=-\frac{2(D+3)}{\k^{2}(4V_{1}+DV^{\ast}_{1})}\,.
  \label{eqn72}
 \end{eqnarray}
 Consequently, the ratios $V_{1}/V^{\ast}_{1}$ and $V_{2}/V^{\ast}_{2}$ 
 of each brane cannot be unity for arbitrary positive $D$, namely, 
 the brane tension becomes anisotropic. 
 From Eq.(\ref{eqn71}), we have
 $V_{1}/V^{\ast}_{1}=V_{2}/V^{\ast}_{2}=(1-k)/(1-l)$, which means that
 the ratio for each brane is the same.
 Furthermore, since $k$ and $l$ cannot be beyond unity,
 the sign of both $V_{1}$ and $V^{\ast}_{1}$ is always
 negative and the relative size between $L$ and $y_{2}$ determines
 the sign of both $V_{2}$ and $V^{\ast}_{2}$. 
 Taking the limit  $L\gg y_{2}$ to be the infinitely fixed observable brane
 \cite{infRS} ,
 both $V_{2}$ and $V^{\ast}_{2}$ approach zero. 
 Moreover, if the extra $D$-dimensional space has infinite dimension,
 taking the limit of $D\rightarrow\infty$, $k\rightarrow\pm 1/2$ and 
 $l\rightarrow 0$, and the ratio becomes
 \begin{eqnarray}
  \frac{V_{1}}{V^{\ast}_{1}}=\frac{V_{2}}{V^{\ast}_{2}}=
  \frac{1}{2}\,,\frac{3}{2}\,.\label{eqn73}
 \end{eqnarray}
 Thus, $V_{1}$ and $V^{\ast}_{1}$ have the same sign
 and $V_{2}$ and $V^{\ast}_{2}$ do also.

 For zero bulk cosmological constant,
 the warped metric functions $a(y)$ and $c(y)$ 
 have the forms of different power laws whose 
 exponents are similar to the constraints appearing in the Kasner solution of 
 higher dimensional cosmology.
 Furthermore, in the case of an infinite orbifold dimension 
 the brane tension of the observable brane becomes zero.
% 
%%%%%%%%%%%%%%%%%%%%%%%%%%%%%%% Summary %%%%%%%%%%%%%%%%%%%%%%%%%%%%%%%%
\section{Summary and Discussion}

 $\hspace{0.5cm}$
 We study the warped metric in the $(5+D)$-dimensional Einstein equation
 with an extra dimension compactified on the  orbifold $S^{1}/Z_{2}$, 
 where two $(3+D)$-branes 
 are fixed on the $y$ direction in orbifold compactification,
 the hidden brane at $y=0$ and the observable brane at $y=L$. 
 It is assumed that the energy-momentum tensor on the brane has
 anisotropic brane tension, anisotropic density, and anisotropic
 pressure. 
 With the ansatz metric in this paper,
 the warped metric function $a(y)$ of four dimensions is generally
 different from that $c(y)$ of $D$ dimensions.
 We solved the Einstein equations in this setup.

 For a negative bulk cosmological constant, 
 whether the integration constant $P_{1}$ in the differential equation 
 coming from the Einstein equation is
 non zero or zero controls the forms of $a(y)$ and $c(y)$.
 If $P_{1}\neq 0$, $a(y)$ and $c(y)$ have
 the forms of different hyperbolic functions, and
 we pointed out that the brane tension becomes anisotropic.
 If $P_{1}=0$, $a(y)$ and $c(y)$ have the same form of exponential
 factor and the brane tension becomes isotropic.
 Furthermore, if the observable brane has negative brane tension,
 the warped metric function is the exponential damping factor and this case
 is similar to five-dimensional Randall-Sundrum scenario.
 For positive bulk cosmological constant, the integration constant
 $P_{1}$ is required to be nonzero in order for the solution
 of the differential equation to exist.
 $a(y)$ and $c(y)$ have the forms of different
 trigonometric functions and the brane tension becomes anisotropic.
 On the other hand, a zero bulk cosmological constant causes
 $a(y)$ and $c(y)$ to have the forms of different power laws whose
 exponents are constrained and thus the brane tension becomes anisotropic. 

 As mentioned above, the dynamics of the differential equation depend on
 the integration constant $P_{1}$ which is determined by the initial
 condition. 
 The mechanism for fixing the value of $P_{1}$ is unknown; however,
 it is expected that this is determined via the dynamics of the underlying
 physics, namely, the initial configuration of the brane world.
 In section 2, we derived the cosmological evolution equation 
 in the isolated two-brane system embedded in $5+D$ dimensions with
 warped metric.
 We are going to study the cosmology constrained by two branes in this setup.
 Moreover, it is necessary to explore the massless gravitational
 fluctuations about our classical solution obtained here and to study 
 the stabilization mechanism for compactification.
 Finally, we expect that this setup may be connected to the 
 $D$-brane configuration in the framework of superstring theory.
%
%%%%%%%%%%%%%%%%%%%%%%%%%%%% Acknowledgements %%%%%%%%%%%%%%%%%%%%%%%%%
%
\section*{Acknowledgments}
 We would like to thank T. Matsuoka (Koggakan University) and M. Matsunaga
 (Mie University) for the hospitality. 
%
%%%%%%%%%%%%%%%%%%%%%%%%%%%% Appendix %%%%%%%%%%%%%%%%%%%%%%%%%
%
\section*{Appendix: Review of the Kasner Solution}
 In this appendix, we review the Kasner solution in higher
 dimensional cosmology \cite{LEcosm,kasner}.
 The original Kasner cosmology is famous as an example of the anisotropic 
 four-dimensional cosmological model, and the metric of Kasner form is 
 \begin{eqnarray}
  ds^{2}=dt^{2}-t^{2p}dx^{2}_{1}-t^{2q}dx^{2}_{2}-t^{2r}dx^{2}_{3}\,,
  \label{eqn74}
 \end{eqnarray}
 where $p$, $q$, and $r$ are parameters.
 The Kasner cosmology corresponds to the vacuum (empty) cosmological model 
 where the numbers $p$, $q$, and $r$ satisfy the constraints 
 \begin{eqnarray}
  p+q+r=1\,,\;\; p^{2}+q^{2}+r^{2}=1\,.\label{eqn75}
 \end{eqnarray}
 The above equations are determined via the Einstein equations.
 The space becomes anisotropic if at least two of the three $p$, $q$, and $r$
 are different.
 
 Next, we describe an extension of the four-dimensional Kasner cosmology to 
 $4+D$ dimensions, and the metric is given by \cite{LEcosm,kasner}
 \begin{eqnarray}
  ds^{2}=dt^{2}-a(t)^{2}d\vec{x}^{2}
        -c(t)^{2}\left(dz^{2}_{1}+\cdots +dz^{2}_{D}\right)\,.
 \label{eqn76}
 \end{eqnarray}
 Here $a(t)$ and $c(t)$ represent the scale factor of the three-space 
 and that of the extra $D$-space, respectively.
 This metric corresponds to $n\equiv 1$, $b\equiv 0$, and $y$ independence of 
 $a$ and $c$ in Eq.(\ref{eqn2}).
 Moreover, there are no contributions of
 the bulk and the brane due to the emptiness.
 We can rewrite the Einstein equations by performing the appropriate 
 linear combinations of Eqs.(\ref{eqn7}), (\ref{eqn8}), and (\ref{eqn10}) 
 \begin{eqnarray}
 && 3\frac{\add}{a}+D\frac{\cdd}{c}=0\,, \nonumber\\
 &&\frac{\add}{a}+2\left(\frac{\ad}{a}\right)^{2}
 +D\frac{\ad}{a}\frac{\cd}{c}=0\,,\nonumber\\
 && \frac{\cdd}{c}+(D-1)\left(\frac{\cd}{c}\right)^{2}
 +3\frac{\ad}{a}\frac{\cd}{c}=0\,.\label{eqn77}
 \end{eqnarray}
 We take the power-law form (so-called Kasner solution) as
 \begin{eqnarray} 
  a(t)&=& a_{0}\left(\frac{t}{t_{0}}\right)^{p}\,,\nonumber\\
  c(t)&=& c_{0}\left(\frac{t}{t_{0}}\right)^{q}\,,\label{eqn78}
 \end{eqnarray}
 where $a_{0}$, $c_{0}$, and $t_{0}$ are constants and 
 the scale factors are normalized to be zero at $t=0$.
 The exponents $p$ and $q$ are subject to the constraints 
 \begin{eqnarray}
  3p+Dq&=&1\,,\nonumber\\
  3p^{2}+Dq^{2}&=&1\,.\label{eqn79}
 \end{eqnarray}
 The above equations can be simply checked by the substitution of 
 Eq.(\ref{eqn78}) into Eq.(\ref{eqn77}).
 Solving this, we have
 \begin{eqnarray}
  p&=&\frac{3\pm\sqrt{3D(D+2)}}{3(D+3)}\,,\nonumber\\
  q&=&\frac{D\mp\sqrt{3D(D+2)}}{D(D+3)}\,.\label{eqn80}
 \end{eqnarray}
 Taking the upper sign in Eq.(\ref{eqn80}), these solutions describe
 the case where the scale factor $a(t)$ of three-dimensional space expands 
 while $c(t)$ of the extra $D$-dimensional space shrinks.

 In the case of zero bulk cosmological constant,
 the form of the metric functions with $y$ dependence
 in Eq.(\ref{eqn66}) resembles the 
 Kasner solutions with $t$ dependence in Eq.(\ref{eqn78}).
 Moreover, the Kasner solutions with radion potential in the framework of
 large extra dimensions are discussed
 in Ref. \cite{LEcosm}.
%
%%%%%%%%%%%%%%%%%%%%%%%%%%%% reference %%%%%%%%%%%%%%%%%%%%%%%%%
%
 
%%%%%%%%%%%%%%%%%%%%%%%%%%%%%%%%%%%%%%%%%%%%%%%%%%%%%%%%%%%%%%%%
\end{document}